\documentclass[12pt,preprint]{aastex}
\begin{document}

\def\ergs{\ifmmode{~{\rm erg~s^{-1}}}\else{~erg~s$^{-1}$}\fi}

\title{Interpreting the Variability of Double-Peaked Emission Lines in Active Galactic Nuclei with Stochastically Perturbed Accretion Disk Models}
\author{H\'el\`ene M.L.G. Flohic, Michael Eracleous}
\affil{Pennsyvania State University}

\begin{abstract}
In an effort to explain the short-timescale variability of the broad, double-peaked profiles of some active galactic nuclei, we constructed stochastically perturbed accretion disk models and calculated H$\alpha$ line profile series as the bright spots rotate, shear and decay. We determined the dependence of the properties of the line profile variability on the spot properties. We compared the variability of the line profile from the models to the observed variability of the H$\alpha$ line of Arp~102B and 3C~390.3. We find that spots need to be concentrated in the outer parts of the line emitting region to reproduce the observed variability properties for Arp~102B. This rules out spot production by star/disk collisions and favors a scenario where the radius of marginal self-gravity is within the line emitting region, creating a sharp increase in the radial spot distribution in the outer parts. In the case of 3C~390.3, all the families of models that we tested can reproduce the observed variability for a suitable choice of model parameters. 

\end{abstract}


\section{Introduction}

\subsection{Background and Motivation}

Double-peaked broad Balmer emission lines are found in 20\% of radio loud active galactic nuclei (AGNs) at $ z<0.4 $ (Eracleous \& Halpern 1994; 2003) and 4\% of the Sloan digital Sky Survey (SDSS) quasars at $z<0.33$ \citep{Sal03}. A number of models have been suggested for the origin of Double-peaked emission lines (DPEL), but most of them face significant challenges when compared to observations. The velocity drift of the two peaks and the single-peaked profile of high-ionization emission lines \citep{Eracal97} are not consistent with the binary black hole scenario suggested by Gaskell (1983). The observed response of the line profiles to changes in the ionizing flux \citep{Dietal98, Obrien98} disagrees with predictions from the bipolar outflow model proposed by Zheng et al. (1990). The disagreement between the observed radio lobe geometry and the outflow geometry required to fit the line profile is also a challenge for this model. Wanders et al. (1995) suggested that the double-peaked profile could be produced by a collection of anisotropically illuminated clouds. Even though this scenario agrees with most observations, it has theoretical difficulties since these clouds would be destroyed rapidly by collisions or drag. Finally, Chen, Halpern, \& Filippenko (1989) and Chen \& Halpern (1989) suggested that the DPEL are emitted in a zone in an accretion disk over a narrow range of radii (i.e. $r_{\rm out}/r_{\rm in}< 10$). In this scenario, the high ionization emission lines are produced at the base of a disk-wind \citep{CS87}. Thus, the profiles of the high-ionization lines are single-peaked because of radiative transfer effects in the accelerating outflow \citep{MC97}. The limitation to this model is that the accretion disk structure required to explain the variability of the line profiles cannot be axisymmetric. However one does not expect a perfectly axisymmetric accretion disk so this is not a major challenge. The model of DPEL production in the accretion is the one that holds up best to close scrutiny, thus we adopt it as our working model. 

DPEL profiles are observed to vary on timescales of months to years, i.e. on timescales of the order of the dynamical time or longer (e.g., Veilleux \& Zheng 1991; Zheng et al. 1991; Marziani et al. 1993; Romano et al 1998; Gilbert et al. 1999; Sergeev et al. 2000; Shapovalova et al. 2001;
Storchi-Bergmann et al. 2003; Gezari et al. 2007). The line profile variability  does not appear to be correlated to changes in the line and/or continuum flux so it likely traces changes in the accretion disk structure. The long-timescale variablity of the DPEL profile of some objects has been successfully modeled by the precession of a non-axisymmetric accretion disk such as an elliptical disk or a disk with a spiral arm \citep[and references therein]{Gezarith, Storchietal03, Shap01, Gilbert}. These models, however, fail to explain the long timescale variability of some objects and the short-timescale variability of all objects \citep{Lewis05}. 

Other attemps at explaining the line profile variability through perturbations of the disk structure have introduced bright spots over an axisymmetric accretion disk. Newman et al. (1997) succesfully modeled the variation of the H$\alpha$ peak intensity ratio of Arp~102B with a single spot rotating within the disk, but Gezari et al. (2007) were not able to apply the same model to the same object at a different time period. It is noteworthy that a similar bright spot model was recently used by Turner et al. (2006) to explain the variability of the Fe K$\alpha$ line profile of Mrk 766 in the X-ray band. Sergeev et al. (2000) tried a different approach; they modeled the variations of Arp~102B with a collection of 1500 clouds in a disk and were able to reproduce the root mean square (r.m.s.) spectrum and a quasi-sinusoidal variation of the excess flux. This model however introduced phase effects and an edge-on disk in disagreement with the inclination obtained from fitting the total line profile \citep{CH89, Chenal89, Gezarith}. A  collection of clouds orbiting the central black hole can also explain the power-law shape of the X-ray power spectra observed in AGNs \citep{Abram91, BO95}. Finally, Pariev \& Bromley (1998) modeled the Fe K$\alpha$ line profile resulting from a turbulent disk, which was created by adding a random frequency shift to each pixel in the accretion disk. Motivated by the ideas of Sergeev et al. (2000), we investigate in this paper whether a stochastically perturbed disk model can explain the short and long timescale variation of the DPEL H$\alpha$ profiles of the two best-monitored objects: Arp~102B and 3C~390.3. 

\subsection{The Physical origin of Bright Spots}

Many processes can lead to density or temperature inhomogeneities, hence brightness perturbations, in the accretion disk: self gravity, disk-star collisions, and baroclinic vorticity. The properties of the inhomogeneities (size, radial distribution, lifetime, resistance to Keplerian shear) depend on their specific formation process, as we outline below. 

\begin{itemize}
\item A disk will fragment under the effect of self-gravity when the Jeans instability sets in. The radius of marginal self-gravity depends on the properties of the disk and the central object and can be as low as 500~$r_{\rm g}$ ($r_{\rm g}\equiv GM/c^2$ is the gravitational radius) for objects with an Eddington ratio of $L/L_{\rm Edd}\sim 10^{-3}$ \citep{GT04}. The radius of marginal self-gravity can be even smaller with more extreme disk properties than those assumed by these authors, such as a Shakura-Sunyaev viscosity parameter less than 0.3 \citep{SS73}. The size of the clumps will be smaller than the local Jeans radius $r_{\rm J}=c_{\rm s}/\sqrt{G\rho}\propto\xi^{-3/20}$, with a typical size of $10~r_{\rm g}$ at 1000~$r_{\rm g}$ for a $10^8~M_{\odot}$ black hole. The clumps produced in this manner do not shear with differential rotation and they have a high density (and hence brightness) that varies very little over time. The radial distribution of the clumps should be quite uniform past the radius of marginal self-gravity.

\item Disk-star collisions are thought to occur as frequently as once per day \citep{Zurekal95}. Such collisions increase the local disk temperature, and hence create a bright spot. These spots will shear with Keplerian rotation and will fade away as the material cools down. The typical size of a spot at the time of formation is at least equal to the size of the star, which is very small in units of $r_{\rm g}$. As shown by Zurek et al. (1995), the radial distribution of the spots created by star collisions will increase with decreasing radius since collisions are more frequent at smaller radii due to the increased Keplerian frequency of the star about the black hole: $\nu_{\rm collison}(r)\propto r^{p-1/2}$ with $0 < p < 1/2$.

\item The radial temperature gradient in the accretion disk combined with the Keplerian differential rotation of the disk material will lead to baroclinic vorticity \citep{PSJ07}. Gas drag will then cause the disk material to spiral to the center of the vortex increasing the local density, hence the brightness, of the vortex \citep{BM05}. The lifetime and resistance of the spots to shearing is still a subject of debate. The simulations of Petersen et al. (2007) create long-lived, non-decaying, non-shearing vortices. Barranco \& Marcus (2005) also find the same long-lived vortices when the vorticity is larger than the shearing; but they also find short-lived, shearing vortices when the vorticity is not high enough or when the gas speed is supersonic. Simulations by Johnson \& Gammie (2005) lead to long-lived, non-shearing vortices that slowly decay as $t^{-1/2}$. All of these models produce structures whose size is comparable to the scale height of the disk, which scales as $r^{9/8}$. The distribution of the spots produced by vorticity is weakly dependent on the radius. 
\end{itemize}

We investigate the effects of these different bright spot properties on the variability of DPEL profiles. In \S 2, we present a method to quantitatively characterize the observed line profile variability and apply it to two AGNs: Arp~102B and 3C~390.3. In \S 3, we describe the mathematical formalism for computing line profiles in our model. In \S 4, we explore the parameter space of the model and compare the results to the observations. In \S 5, we discuss the results and their implications . 

\section{Charaterizing the Observed Line Profile Variability}

\subsection{Available Data}

We use the spectroscopic observations of Arp~102B and 3C~390.3 described in Gezari et al. (2007). We chose these two broad-line radio galaxies because they were regularly monitored for extended periods of time and because their black hole masses are accurately measured \citep{LE06}. The observations of Arp~102B were made from June 1983 to September 2004, while the observations of 3C~390.3 span a shorter time period, from July 1988 to September 2004. These spectra have moderate resolution (4--6~\AA~or 180--260~km~s$^{-1}$) and cover at least the wavelength range of the H$\alpha$ line.

After the initial processing of the data, the continuum and narrow emission lines were subtracted as described by Gezari et al. (2007). We excluded spectra for which Gezari et al. did not achieve reliable subtraction of the narrow lines. We thus have 91 line profiles with an average time separation of 86 days for Arp~102B and 35 spectra every 174 days, on average, for 3C~390.3. Hence the dynamical timescale of the disk (as calculated in \S \ref{sec:application}) is well sampled. The average spectra of the two objects are shown in Figure \ref{fig:fig_av} with the best fitting axisymmetric disk model superposed (see \S \ref{sec:application}).

All these line profiles were normalized to a constant integrated flux. This was done because the variability of the integrated emission line flux is correlated with the intensity of the ionizing continuum rather than with structural changes in the disk \citep[e.g.][]{Rosenetal92, Kassebaum97, WP97}. We then measure the seven line paramaters described by Gezari et al. (2007): the red and blue peak velocities (in km~s$^{-1}$), the blue peak to red peak flux ratio, the full width at half maximum and at quarter maximum (FWHM and FWQM, in km~s$^{-1}$) and the line displacement from its rest wavelength at half and quarter maximum (in km~s$^{-1}$) [see Figures 3 and 5 of Gezari et al. (2007) for the parameter time series of Arp~102B and 3C~390.3 respectively]. We chose to use these line parameters since they are independent of the normalization of the line profile and do not require any fitting of the line profile by a specific physical model.

\subsection{Method for Characterizing the Variability}

We cannot fit series of model line profiles to the observed data because of the inherently stochastic nature of the model (described in \S \ref{sec:model1}). Comparing the variation of the line profile parameters as a function of time is not meaningful either since a change in the initial observing time or initial conditions will modify the exact values of these parameters. Hence we developed a comparison method based on power spectra, which captures the statistical properties of the variability.


Throughout this paper we use the Lomb-Scargle algorithm (Scargle 1982; Press \& Rybicki 1989) to compute periodograms because our time series are unevenly sampled in time. We tested the robustness of the algorithm by comparing the periodograms of evenly- and unevenly-sampled, simulated time series. While the periodogram of an unevenly-sampled time series might have missing power peaks when compared to the results from an evenly-sampled series, it never has false positive power peaks. The sampling interval has a greater impact on the resulting power-law index of the periodogram: irregularly sampled time series produce significantly flatter periodograms because the Lomb-Scargle algorithm produces inherently noisier periodograms at high frequency than the fast Fourier transform of an evenly sampled time series. To bypass this effect, we sample our simulated times series at the same times as the data.

From the line profile series, we created light curves in narrow velocity intervals ($\delta\lambda=20$~\AA, corresponding to $\Delta v = 914$~km s$^{-1}$) and computed the periodograms. Thus, we produced 2-D periodograms, which are displayed in Figures \ref{fig:2dfft-arp102b} and \ref{fig:2dfft-3c390.3}, for Arp~102B and 3C~390.3 respectively. The probability that a power peak of height $z$ is created at random is then $P\approx N e^{-z}$ where $N$ is the number of frequency bins (110 and 36 for Arp~102B and 3C390.3 respectively). Each 2-D periodogram has 40 velocity bins so one can expect the power to be above the 98\% significance level in up to 88 bins  for Arp~102B and 29 bins for 3C~390.3. The actual number of bins with power above the 98\% significance level is 59 for Arp~102B and 10 for 3C~390.3, which is below the random occurrence expectation.  Moreover these power peaks are mostly located at low frequency where the error bars are large, making the significance of these peaks low. Because the significance of discrete power peaks is low, we concentrate on another aspect of the periodograms, namely the power-law index. The bottom panels of Figures \ref{fig:2dfft-arp102b} and \ref{fig:2dfft-3c390.3} show the collapsed (i.e., summed) periodograms as a function of frequency for all velocities $v<0$ (dashed line) and $v>0$ (continuous line). The power-laws that best fit each periodogram (assuming $z\propto f^\alpha$) are also displayed. For Arp~102B, the power-law of the periodogram with $v<0$ has an index of $-0.60\pm 0.06$ and that with $v>0$, $-0.47\pm0.06$. For 3C~390.3, the $v<0$ periodogram has a power-law index of $-0.6\pm0.1$ and the $v>0$ periodogram, $-0.6\pm0.1$. This indicates that the profile variability of Arp~102B is somewhat more asymmetric than that of 3C~390.3. The right panel shows the total power per unit frequency in each velocity interval. The velocity intervals with the highest average power are the most variable parts of the line profile.

Another way to characterize the line profile variability is through the periodograms of the times series of profile parameters. From the line parameters that we previously measured, we construct time series, and calculate the corresponding periodograms. The line profile parameter periodgrams for Arp~102B and 3C~390.3 are displayed in Figures \ref{fig:lineparam-arp102b} and  \ref{fig:lineparam-3c390.3} respectively. Power peaks seen in these periodograms are not considered significant due to their low power and their number which is consistent with chance occurrence. 
The power-law indices of all the line parameter periodograms of Arp~102B are negative and vary between $-0.30$ and $-0.85$. 
The line parameter periodograms of 3C~390.3 are flatter than those of Arp~102B, with power-law indices varying between $-0.76$ and 0. The values of all these indices are given in the captions of Figure \ref{fig:lineparam-arp102b} and \ref{fig:lineparam-3c390.3}. The error bars on these power-law indices are $\sim 0.1$, which means that: (a) the power-law indices of the periodograms of the red peak position and of the FWQM are significantly different from each other, (b) the power-law indices of the blue peak position periodogram is marginally different between the two objects, and (c) all the other power-law indices are consistent with each other. 

We also calculated the root mean square (rms) amplitudes of the line profile parameters for each object (listed in Table \ref{tab:rms}). The fractional rms amplitudes (rms amplitude divided by mean) of the line parameters give an alternative estimate of the amplitude of the variability. The fractional rms amplitude of the red and blue peaks positions and displacement at half and quarter maximum of 3C~390.3 are much larger than those of Arp~102B. This means that the line profile changes of 3C~390.3 have a much larger amplitude than Arp~102B.

Hence the line profile variability properties of Arp~102B and 3C~390.3 are quite different from one another whether we consider the 2-D periodograms, the line parameter periodograms or the fractional rms amplitude. With this in mind, we tested accretion disk structure models and their induced variability characteristics in search of a model that could reproduce the observed line profile variability of these two AGNs.

\section{Description of the Models\label{sec:model1}}

We use the formalism laid out by Chen et al. (1989) and Chen \& Halpern (1989) to calculate the double-peaked emission line profile from an accretion disk in the weak-field limit. We compute the line profile numerically by evaluating 

\begin{equation}
f_{\nu} \propto\int\int\xi~I_{\nu_{\rm e}}(\xi ,\phi, \nu)~D^3(\xi , \phi )~\Psi (\xi , \phi )~d\phi~d\xi 
\end{equation}
over the surface of the disk. Here, $\phi$ is the azimuthal angle in the disk frame and $\xi$ is the distance from the central black hole in units of $r_{\rm g}$. The velocity structure of the disk is contained in the Doppler factor $D(\xi , \phi )$ and the gravitational redshift and light bending effects are included in $\Psi (\xi , \phi )$. These factors were computed by Chen et al. (1989) for a circular disk and depend on the inclination $i$ of the axis of the disk to the line of sight\footnote{The notation $\psi(\xi,\phi)$  was introduced by Eracleous et al. (1995). This corresponds to the function $g(D)$ used by Chen et al. (1989)}. 

The specific intensity $I_{\nu_{\rm e}}(\xi ,\phi ,\nu_e )$ contains information on the local line profile (assumed to be Gaussian) and the disk emissivity function $\epsilon (\xi ,\phi )$:

\begin{equation}
I_{\nu_{\rm e}}(\xi ,\phi, \nu_e) \propto \epsilon (\xi ,\phi )\exp\left[-\frac{(\nu_{\rm e}-\nu_0)^2}{2\sigma^2}\right]
\end{equation}
where $\sigma$ is the line broadening parameter, $\nu_{\rm e}$ is the emitted frequency of the photon and  $\nu_0$ is the rest frequency of the emitted photon. 

The stochastically perturbed model was implemented by creating the appropriate initial disk emissivity function $\epsilon (\xi ,\phi )$. Over an underlying power-law emissivity of index $-q$ ($q>0$), we add $n$ circular spots. Each spot has a random position in the disk $(\xi_{\rm i}, \phi_{\rm i})$, a radial Gaussian profile of spread $w_{\rm i}$, and a contrast $C_{\rm i}$:

\begin{equation}
\epsilon (\xi ,\phi, t_0 )\propto\xi^{-q}\left\{1+\sum_{i=1}^n C_{\rm i}\exp \left[-\frac{\xi_{\rm i}^2(\phi-\phi_{\rm i})^2}{2w_{\rm i}^2}\right] \exp\left[-\frac{(\xi-\xi_{\rm i})^2}{2w_{\rm i}^2}\right]\right\}
\end{equation}

The values of $\xi_{\rm i}$ and $\phi_{\rm i}$ are randomly generated. The radial distribution of the spots can be either uniform or a power-law, while their azimuthal distribution is always uniform. The spot size, $w_{\rm i}$, scales as the disk scale height ($\propto \xi^{9/8}$) and the spot size at the inner radius ($w_{\rm min}$) is an adjustable parameter. The contrast, $C_{\rm i}$, is randomly generated from a Gaussian distribution with a standard deviation of 1; the average value of $C_{\rm i}$ is an adjustable parameter. The average number of spots, $n$, is kept constant through a given simulation.

\begin{figure}
\centerline{\includegraphics[width=2.7in]{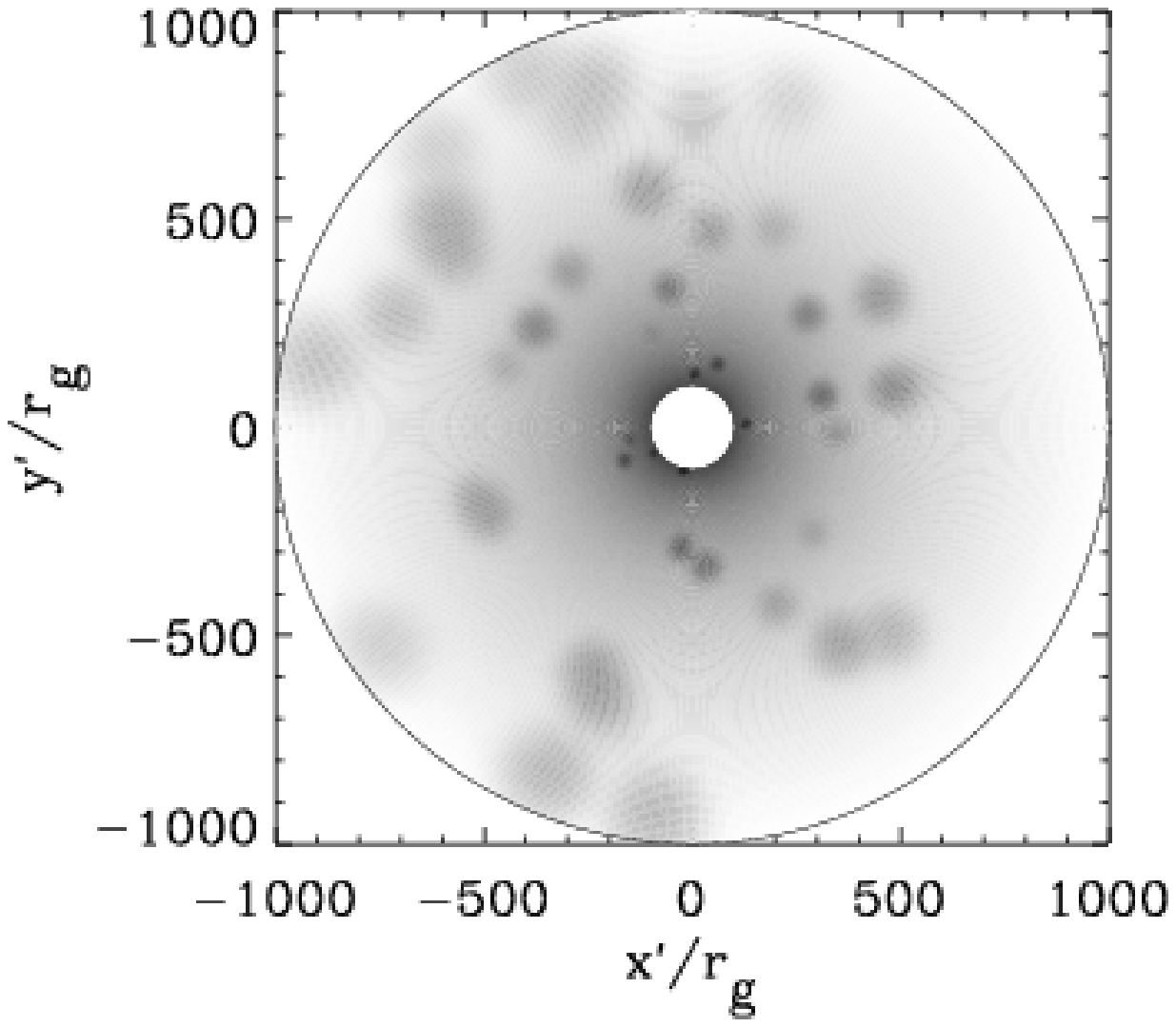}}
\centerline{\includegraphics[width=2.7in]{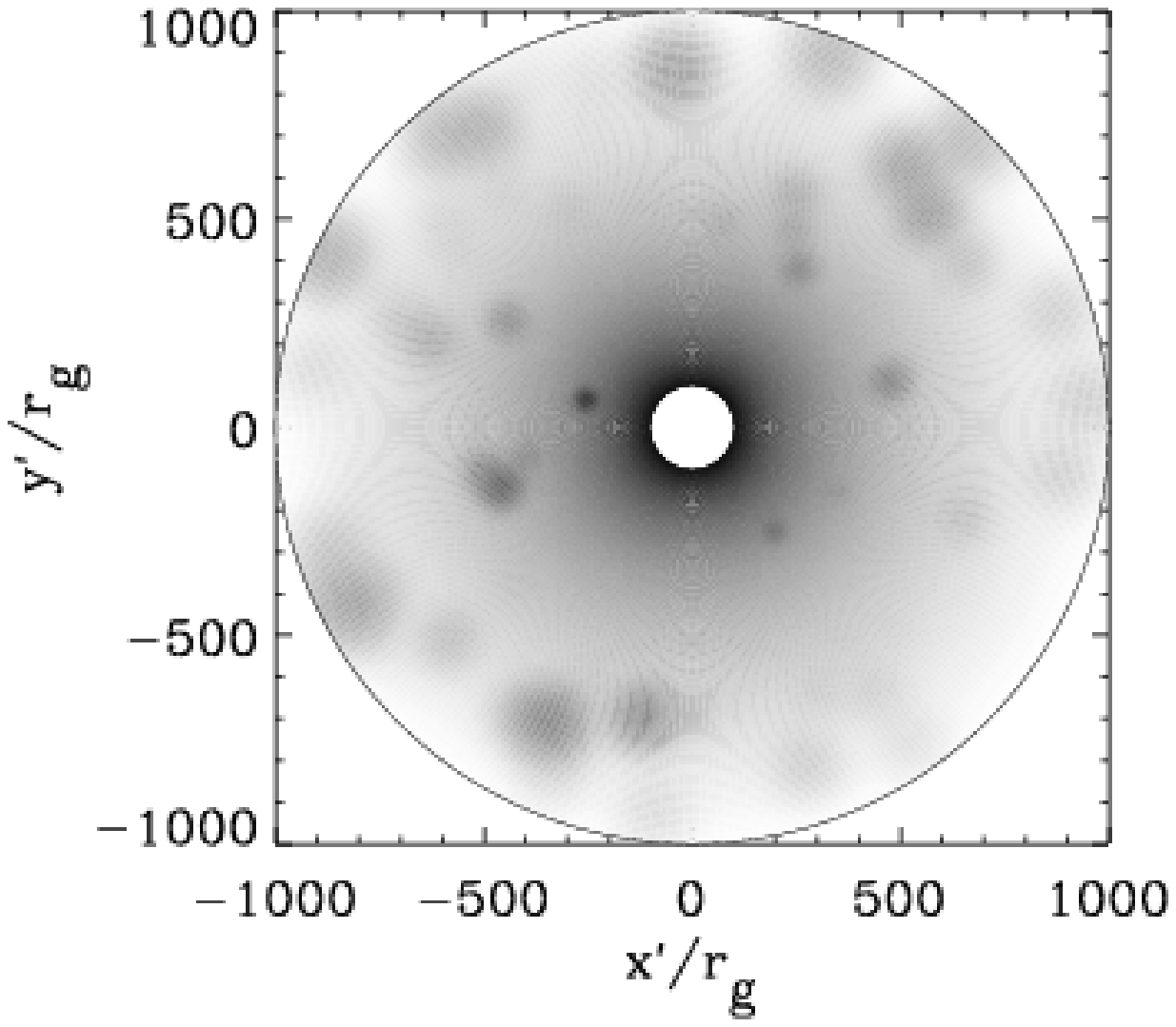}}
\centerline{\includegraphics[width=2.7in]{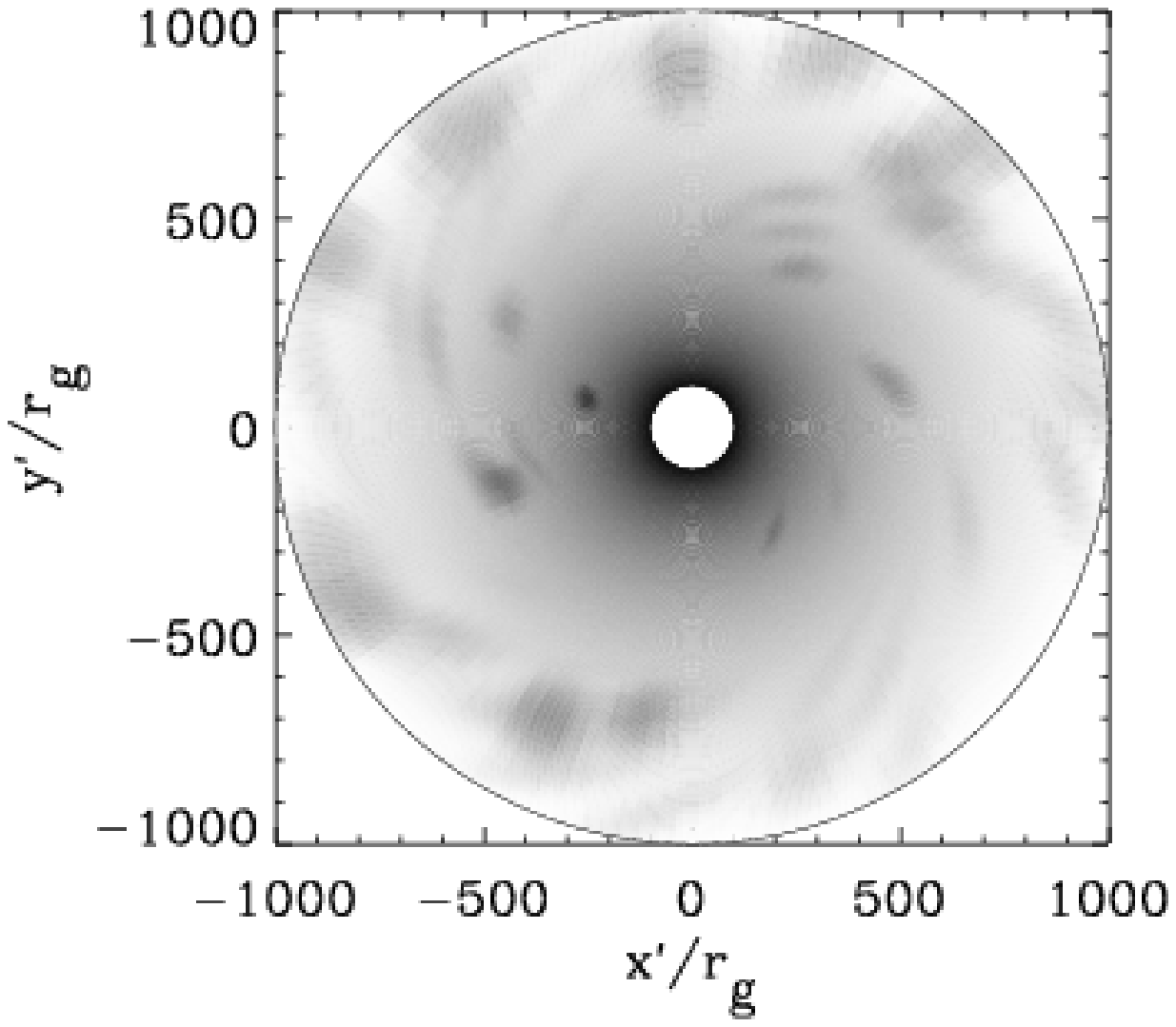}}
\caption{Snapshots of the disk emissivity function for the stochastically perturbed disk model without decay or shearing of the spots (top), with decay but without shearing (middle), and with decay and shearing (bottom). Each simulation was started with identical initial conditions and the snapshots were all made 6 months after the initial time. Darker shades represent areas of higher emissivity. 
\label{fig:snapshot}}
\end{figure}

\begin{figure}
\centerline{\includegraphics[width=6in]{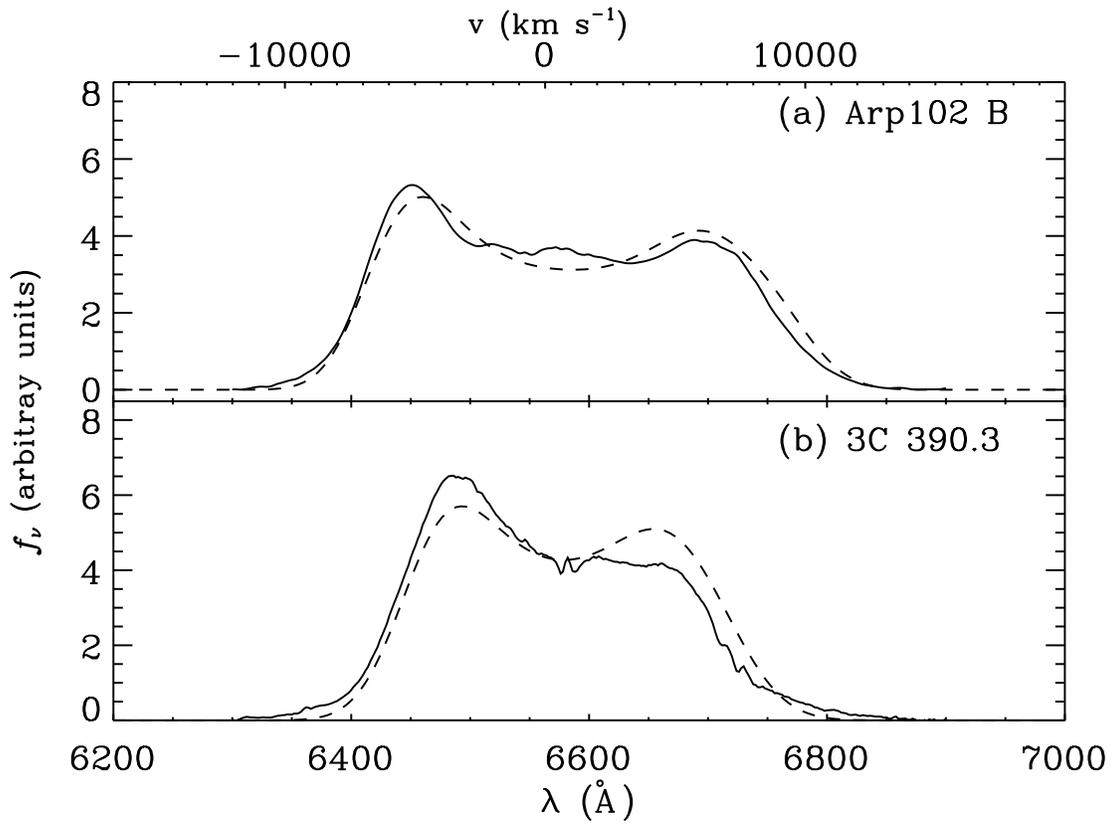}}
\caption{Average spectra of Arp~102B and 3C~390.3 with the best fitting axisymetric disk model superposed as a dashed line. The model is described in \S \ref{sec:model1} and the values of the model parameters are given in sections \S \ref{sec:app-arp102b} and \ref{sec:app-3c390.3}.
\label{fig:fig_av}}
\end{figure}

\begin{figure}
\centerline{\includegraphics[width=6in]{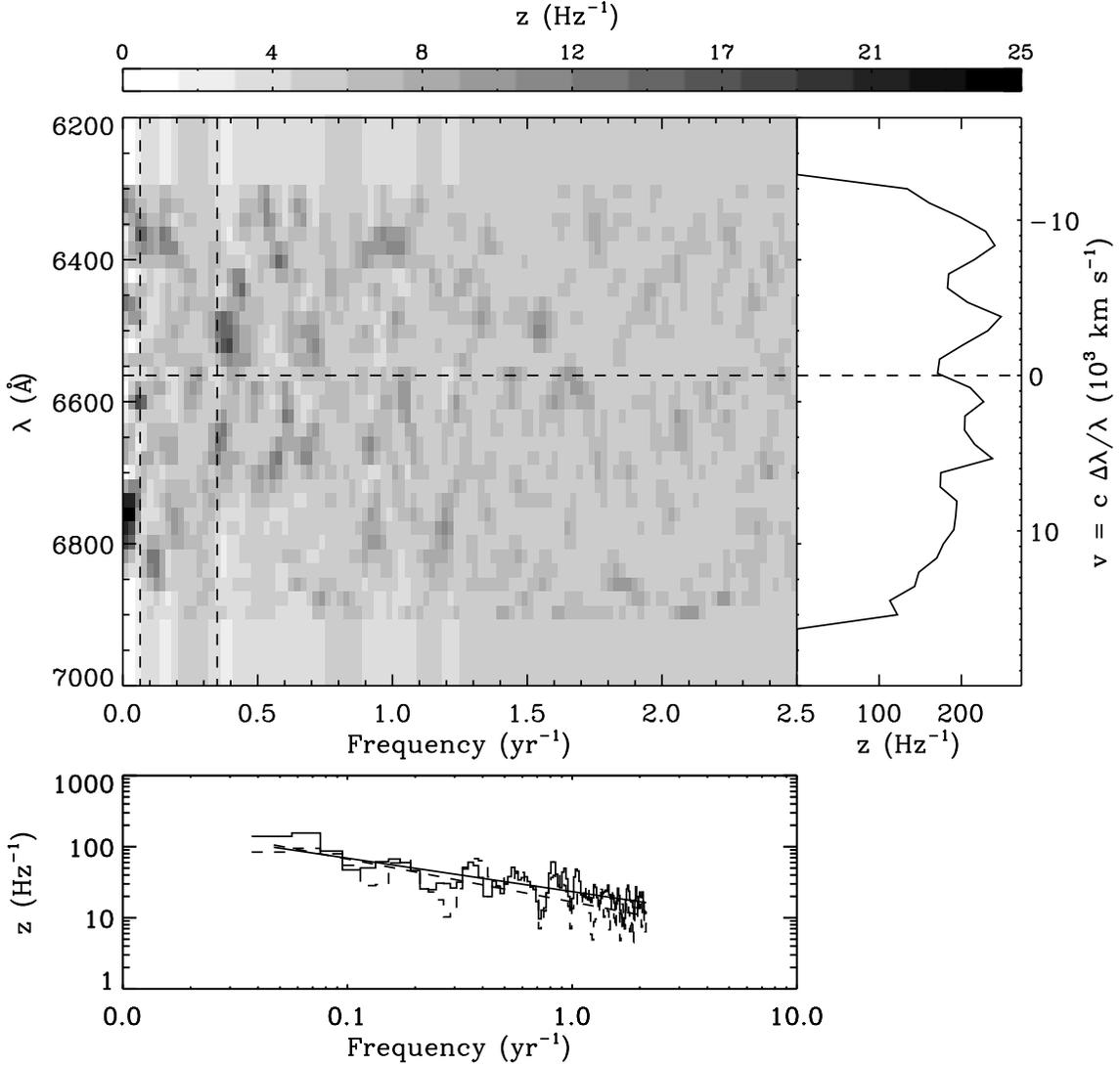}}
\caption{2D periodogram of Arp~102B computed in 20~\AA~bins. The grayscale indicates the power. The horizontal dashed line represents the rest wavelength of the emission line. The vertical dashed line represents the Keplerian frequency of the outer radius of the line emitting region. The total power per unit frequency in each velocity bin is displayed in the right panel. The bottom panel shows the total power per unit frequency as a function of frequency, summed over all bins with $v>0$ (continuous) and $v<0$ (dashed), along with the best fitting power-laws (assuming $z\propto f^\alpha$). The indices of these power-laws are $-0.47\pm0.06$ and $-0.6\pm0.06$, respectively. (Note that the frequency has a logarithmic scale in the bottom panel and a linear scale in the top panel.)\label{fig:2dfft-arp102b}}
\end{figure}

\begin{figure}
\centerline{\includegraphics[width=6in]{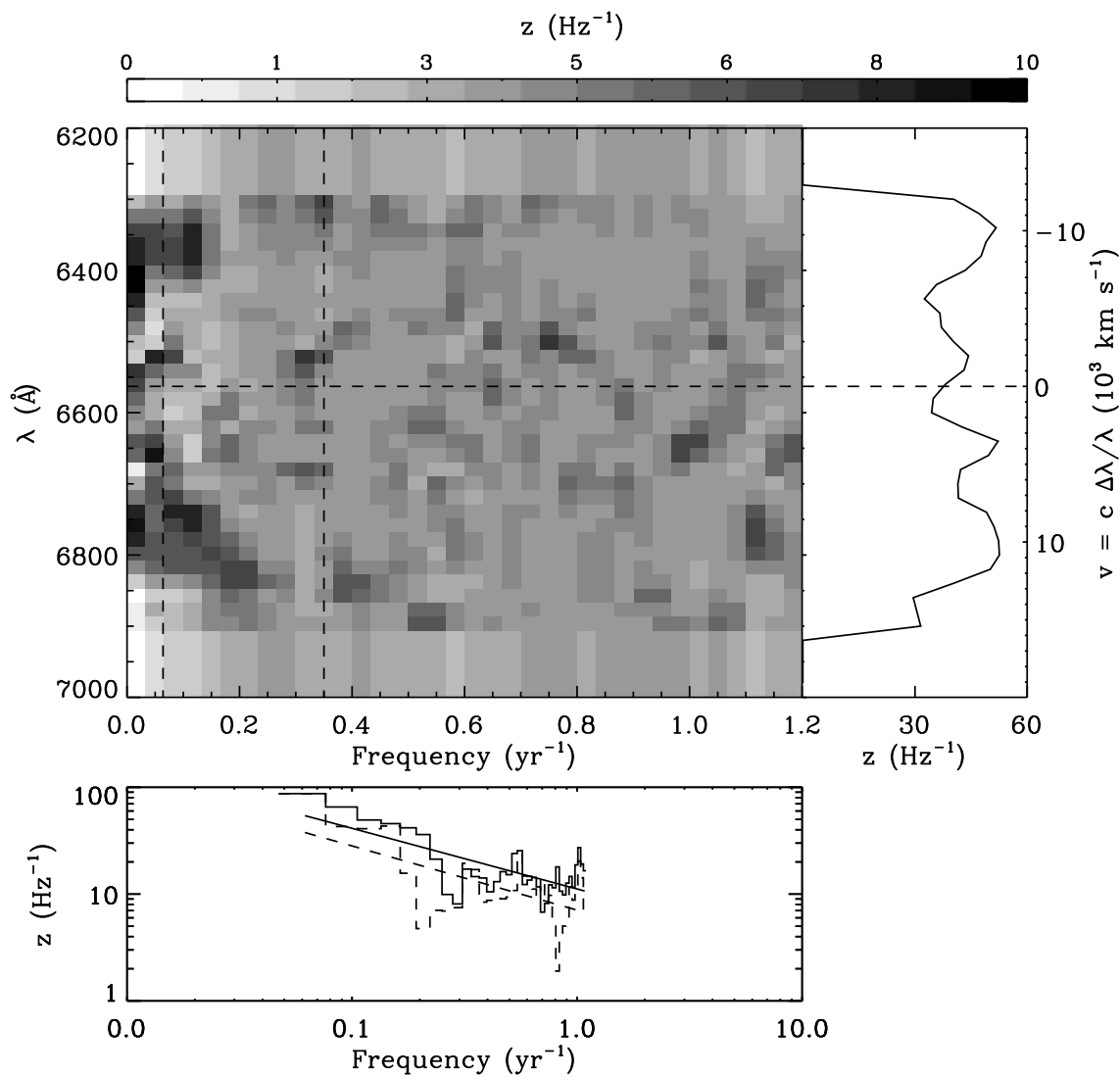}}
\caption{Same as Figure \ref{fig:2dfft-arp102b}, but for 3C~390.3. The left and right vertical dashed lines in the top panel represent the Keplerian frequencies of the outer and inner radii of the line-emitting part of the disk, respectively. The power-law indices are $-0.6\pm0.1$ and $-0.6\pm0.1$ for $v<0$ and $v>0$, respectively. (Note that the frequency has a logarithmic scale in the bottom panel and a linear scale in the top panel.) \label{fig:2dfft-3c390.3}}
\end{figure}

\begin{figure}
\centerline{\includegraphics[width=6in]{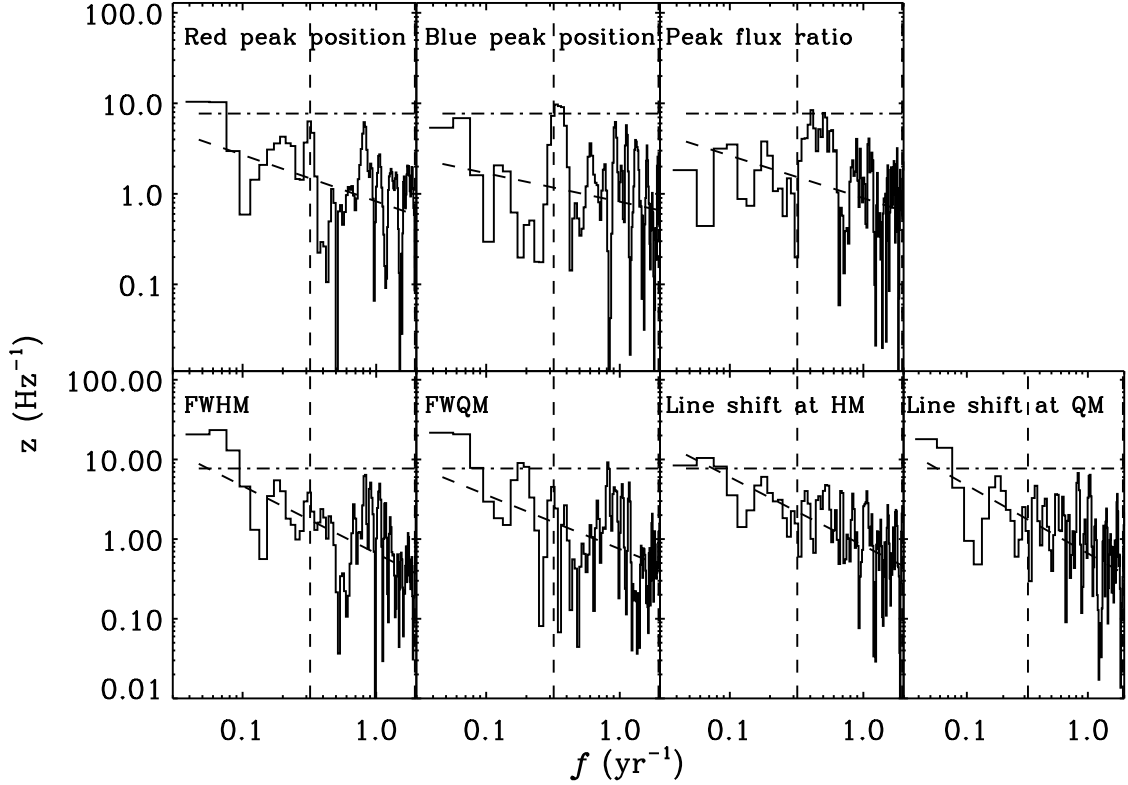}}
\caption{Periodogram of the line profile parameters of Arp~102B. The horizontal dash-dot line indicates the 98\% signifance limit of the power peaks. The slanted dashed line shows the best-fitting power-law (assuming $z\propto f^\alpha$). The power-law indices are (left to right, then top to bottom) $-0.50$, $-0.30$, $-0.45$, $-0.80$, $-0.70$, $-0.85$, $-0.85$ with an uncertainty of $\sim0.1$. The vertical dashed line indicates the Keplerian frequency of the outer radius of the line-emitting part of the disk, obtained from the model fits described in \S \ref{sec:application}. \label{fig:lineparam-arp102b}}
\end{figure}

\begin{figure}
\centerline{\includegraphics[width=6in]{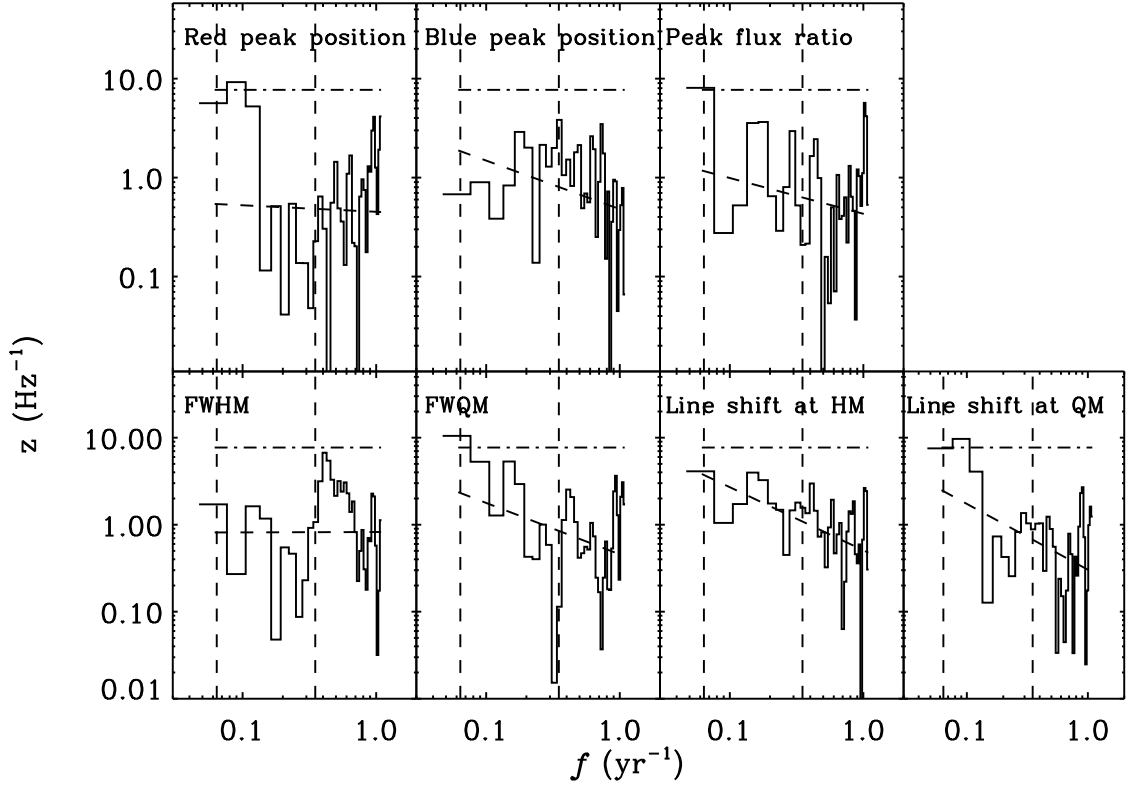}}
\caption{Same as Figure \ref{fig:lineparam-arp102b}, but for 3C~390.3. The left and right vertical dashed lines represent the Keplerian frequencies of the outer and inner radii of the line-emitting part of the disk, respectively. The power-law indices are (left to right, then top to bottom) $-0.07$, $-0.49$, $-0.36$, $0.00$, $-0.59$, $-0.73$ and $-0.76$  with an uncertainty of $\sim0.1$.\label{fig:lineparam-3c390.3}
}
\end{figure}

We created three families of models from the basic spot distribution corresponding to the physical models discussed in \S 1. The properties of these families are as follows:

\begin{enumerate}
\item The spots do not shear and do not decay as their centers rotate differentially around the central black hole. This model is relevant to disks in which self gravity is important or disks with long-lived non-shearing vortices, such as those described by Barranco \& Marcus (2005) and Petersen et al. (2007). A snapshot of the disk as described by this model is shown in Figure \ref{fig:snapshot}a.
\item The spots do not shear, but they do decay with time as their centers rotate differentially, which is relevant to the Johnson \& Gammie (2005) model. A snapshot of the disk as described by this model is shown in Figure \ref{fig:snapshot}b.
\item The spots shear due to the Keplerian rotation of the disk and also decay. This model is relevant to spots created by disk-star collisions (power-law radial distribution of spots) and short-lived spots created by weak vorticity in the Barranco \& Marcus (2004) model (with a uniform radial distribution of spots). A snapshot of the disk as described by this model is shown in Figure \ref{fig:snapshot}c.
\end{enumerate}

The Keplerian timescale for the rotation of the spots in the accretion disk depends on the mass of the black hole, $M_{\rm BH}$, which can be a free parameter, in principle. The fading of the spots was treated by multiplying the initial value of $C_{\rm i}$ by an exponential decay function whose decay time $\tau_{\rm decay}$ is a free parameter. This decay time is assumed to be a function of radius ($\propto \xi^{3/2}$), i.e. proportional to the local dynamical time. When the spots decay away, new spots have to be created to keep the average number of spots constant. This is done by balancing the average decay rate and the appearance rate.

Hence the free parameters are: $q$, $M_{\rm BH}$, $\xi_{\rm min}$, $\xi_{\rm max}$, $i$, $\sigma$, $\langle C_{\rm i}\rangle$, $w_{\rm min}$, $n$, and $\tau_{\rm decay}$, if the spots decay. The black hole masses are set equal to the values reported by Lewis \& Eracleous (2006) and we assume that $q=3$, following Dumont \& Collin-Souffrin (1990). Over the total baseline spanned by the observations, the inner and outer radius of the line-emitting portion of the accretion disk could vary. For example, the inner radius could increase at a time of increased luminosity of the central engine due to ionization of the inner parts of the accretion disk \citep{CSD90, Shap01, SBal95}. If we let $\xi_{\rm min}$ and $\xi_{\rm max}$ be free parameters for each profile of a time series, the total number of free parameters increases greatly (e.g., 185 or 186 free parameters for Arp~102B). Instead we use an average value of $\xi_{\rm min}$ and $\xi_{\rm max}$ obtained from fitting an axisymmetric model to the average and minimum spectra. The minimum spectrum was constructed by taking the minimum value of the normalized flux from the entire time series at each wavelength. Keeping the radii fixed is justified by the fact that large-amplitude continuum variability (changes by more than an order of magnitude) occurs on timescales considerably longer than the Keplerian timescale at these radii. Therefore noticeable fluctuations in the radii should also occur on timescales considerably longer than the Keplerian timescale at these radii. This reduces the total number of free parameters to 3 or 4. By keeping the radii fixed we are able to isolate the effect of emissivity perturbations on the variability of the line profiles.

The computational implementation of the model requires dividing the disk into a logarithmic grid in $\xi$ and linear grid in $\phi$. We used a $200\times 200$ grid in order to resolve each spot into more than a few pixels and to resolve the effect of Keplerian shear. We chose a logarithmic division of the grid in the radial direction because the bulk of the line flux is produced at small radii (due to the $\xi^{-3}$ emissivity function) and because most of the quantities that depend on radius are described by power laws.

\section{Application \label{sec:application}}

\subsection{Arp~102B\label{sec:app-arp102b}}

We constructed the minimum and average spectra from the 91 line profiles of Arp~102B and fitted both of them with a model employing an axisymmetric power-law emissivity of the form $\epsilon\propto \xi^{-q}$ with $q=3$ and $i=30^\circ$. The best fit results for the average profile are $\xi_{\rm min}=290\pm 30$, $\xi_{\rm max}=1000^{+350}_{-200}$ and $\sigma =1300\pm 400$~km~s$^{-1}$. The best fit parameters to the minimum spectrum are $\xi_{\rm min}=340\pm 40$, $\xi_{\rm max}=940^{+400}_{-200}$ and $\sigma =1200\pm 400$~km~s$^{-1}$, which are quite consistent with the results of the best fit to the average spectrum and with fit results by previous studies \citep{CH89, Chenal89, Gezarith}. The average spectrum and the best fit are shown in Figure \ref{fig:fig_av}a. 

For the analysis of Arp~102B we set the underlying disk parameters to the following values: $\xi_{\rm min}=300$, $\xi_{\rm max}=950$ and $\sigma=1200$~km~s$^{-1}$. The mass of the black hole is set to $10^8~M_{\odot}$ as measured by Lewis \& Eracleous (2006). Hence the dynamical times ($\tau_{\rm dyn}$) at the inner and outer disk radii are 0.08 and 0.45~years, respectively. 

We note that the fit to the average spectrum is not as good as what can be obtained for individual spectra \citep[e.g.,][]{CH89}. We investigated the effect of setting the disk parameters to different values on the resulting variability properties. We used the best fit values found by Chen \& Halpern (1989), namely $i=32^\circ$, $\xi_{\rm min}=350$, $\xi_{\rm out}=1000$ and $\sigma=850$~km~s$^{-1}$ and the values we decided upon in two runs of the first family of models with identical spot properties. We then computed the line parameter periodograms and found that the difference in power-law indices between the two runs was very small $\Delta\alpha=0.08\pm0.04$. This is a result of the subtraction of the average profile before computing the periodogram. Since we are concerned with perturbations about the mean, the exact values of the disk parameters have a negligible effect on our conclusions.

\subsubsection{First Family of Models: Non-Shearing, Non-Decaying Spots}

\begin{figure}
\centerline{\includegraphics[width=6in]{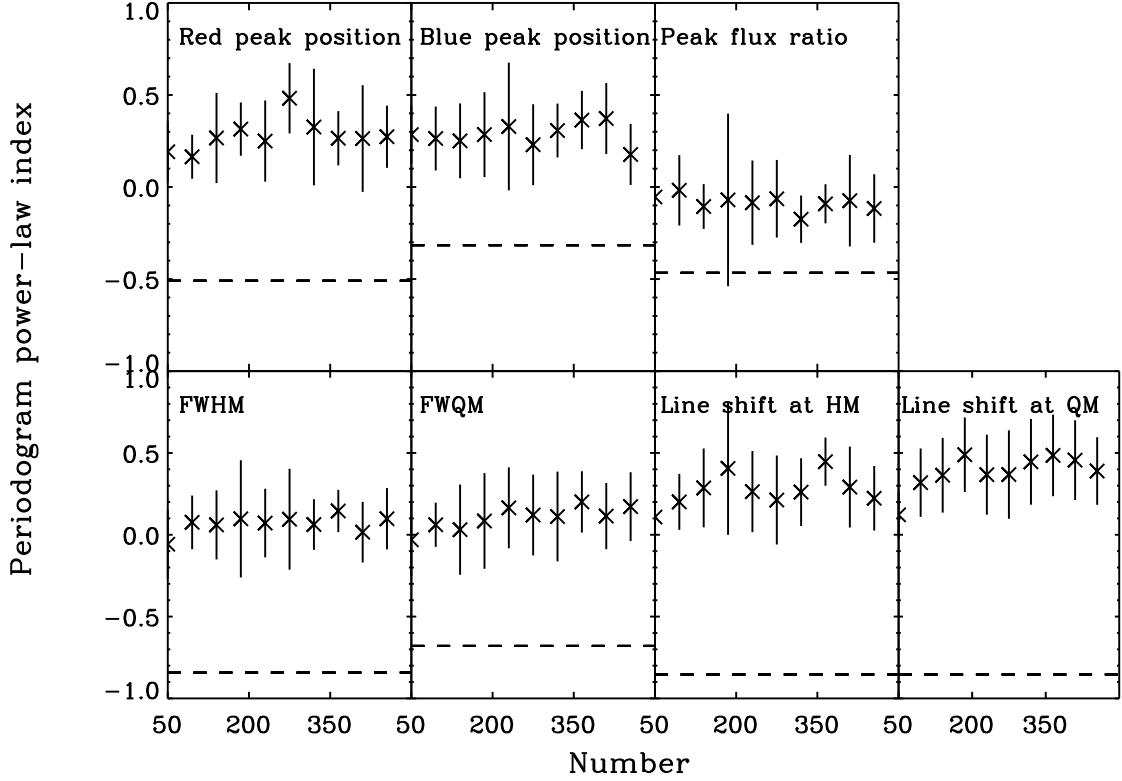}}
\caption{Average and standard deviation of the power-law indices of the line parameter periodograms as a function of the number of spots in the disk. The vertical lines represent the scatter in the power-law indices of the simulations, not the error bars of the measured power-law indices. The model used here includes non-decaying, non-shearing spots. We investigate the effect of varying the number of spots between 50 and 500, while keeping all other parameters constant. The horizontal dashed lines represent the value of the power-law index found in the corresponding observed periodogram of Arp~102B.\label{fig:bin-index-clumpy-nb}
}
\end{figure}

We first characterize the variability of the model with non-shearing, non-decaying spots (i.e. the first family of models). In order to investigate the effects of varying the free parameters, we vary only one parameter at a time (namely, the number, size, contrast, and radial distribution). We then create 100 time series with randomly selected values of the parameter of interest, measure the power-law indices of the line parameter periodograms for each run, and plot the average and standard deviation of the power-law indices against the free parameter value. Figure \ref{fig:bin-index-clumpy-nb} shows as an example the dependence of the power-law indices of the line parameter periodograms on the number of spots with the other spot parameters fixed to $w_{\rm min}=10~r_{\rm g}$, $\langle C_{\rm i}\rangle=10$ and a constant number of spots per unit radius. We calculated the Pearson coefficient and the significance of the Spearman rank coefficient in order to search for any correlation between the parameter of interest and the power-law indices. Since the dispersion of  the power-law indices is large, we set stringent limits for considering these coefficients to be significant ($\|Pe\|>0.8$ and $P_{\rm s}<0.02$). We also examined the plots to look for a break in the correlation and recalculate the coefficients on both sides of the break when we suspect one. Varying the number of spots produces power-law indices consistent with 0 or greater, which does not match the observed values ($-0.30$ and $-0.85$). There is no trend between any power-law index and the number of spots in this model. Using the same approach for the other model parameters of interest, we find that the logarithmic slope of the power spectra depends only weakly, if at all, on the model parameters. The weak trends that we notice are:
\begin{itemize}
\item All the periodograms become steeper as the spot contrast increases.
\item Most of the periodograms become steeper when the spots are concentrated in the outer parts of the disk.
\item The power-law index of the periodogram of line shifts at QM has a broad maximum at a spot size $\sim 10~r_{\rm g}$; it is less than 0.3 for small ($<5~r_{\rm g}$) and large ($>12~r_{\rm g}$) spots while it is $\sim 0.4$ around 10~$r_{\rm g}$.
\end{itemize}

Thus bright spots with $w_{\rm min} < 5~r_{\rm g}$ or $w_{\rm min} > 12~r_{\rm g}$, concentrated in the outer parts of the disk produce the steepest possible power spectra. However, even the steepest model power spectra are too flat to explain the power spectra observed in Arp~102B (see illustration in Fig. \ref{fig:bin-index-clumpy-nb}).


We created 2-D periodograms for each run of the model with non-shearing, non-decaying spots and calculated the power-law indices of the collapsed periodograms for $v<0$ and $v>0$. The difference between the two power-law indices vary from $-0.2$ to $0.2$. The average value for the power-law indices is $-0.2$, which is higher than the observed values for Arp~102B. The power-law indices have a slight dependence on the radial spot distribution: the most negative power-law indices are produced when spots are concentrated in the inner part of the disk. 

We also investigated the effects of varying the model parameters on the fractional rms amplitude. We find that the fractional rms amplitude increases with increasing contrast and size of the spots, but does not depend significantly on the number of spots or the radial distribution. Furthermore, in order to reproduce the observed fractional rms amplitude of Arp~102B, a combination of particularly large size and contrast is necessary ($C_{\rm i} > 20$ and $w_{\rm min} > 10$~r$_{\rm g}$). 

In conclusion, the first family of models can produce asymmetric 2-D periodograms, as observed. However, this family also produces power-law indices that are too high compared to what is observed in Arp~102B (i.e. the simulated power spectra are too flat). 
The agreement between this family of models and the observed properties of the line profile parameters of Arp~102B is best when the spots are concentrated in the outer parts of the disk. But the best agreement with the collapsed periodograms is obtained when spots are concentrated in the inner parts of the disk. 

\subsubsection{Second Family of Models: Shearing, Non-Decaying Spots}

Similarly, we characterize the variability of models whose spots decay and do not shear (i.e., the second family of models) by letting only one parameter free to vary and by measuring the spread in power-law indices in each parameter bin. We find that:
\begin{itemize}
\item The periodograms of the line shifts at QM and HM and of the red peak position steepen when the decay time gets shorter. However the power-law indices of the FWHM and FWQM steepen when the cooling time gets longer.
\item The periodograms of the blue peak position steepen with decreasing initial contrast of the spots.
\item The periodograms steepen when the radial distribution of the spots favors more spots in the outer part of the disk. This correlation is not as strong as for the model without spot decay (first family of models).
\end{itemize}

Compared to the first family of models, the second family produces periodograms whose power-law indices are steeper by 0.2 for a decay time of 5$\tau_{\rm dyn}$ and 0.4 when $\tau_{\rm decay}=\tau_{\rm dyn}$. This brings the simulated power-law indices closer to the observed values. The power-law indices of the periodograms of the peak flux ratio, blue peak position, the FWHM and the FWQM are in good agreement with the observed values of Arp~102B for a $\xi^2$ radial distribution of spots and $\tau_{\rm decay} > 5 \tau_{\rm dyn}$, but the power-law indices for the other line profile parameters are too high by 0.3 to 0.6


The difference between the two power-law indices of the collapsed line flux periodograms varies between $-0.2$ and $0.3$. The average value for the power-law indices is 0, which is greater than for the first family of models and the observed values. The power-law indices show the same dependence on radial spot distribution. The fractional rms amplitude is similar to that of the first family and the decay time has no effect on the fractional rms amplitude. 

\subsubsection{Third Family of Models: Shearing, Decaying Spots}

Finally, for the model whose spots shear and decay (i.e. the third family of models), we find:
\begin{itemize}
\item The periodograms of the FWHM and FWQM steepen with increasing decay time, increasing contrast, and decreasing number of spots.
\item The periodograms of the red and blue peak position steepen with decreasing contrast.
\item The periodograms steepen when the radial distribution of the spots favors more spots in the outer part of the disk. The strength of this correlation is comparable to that observed for the second family of models.
\end{itemize}

This family of models produces periodograms with power-law indices that are exactly like those of the second family of models, within the dispersion. The 2-D periodograms produced by this family of models have the same behavior as those of the second family of models. The fractional rms amplitude is similar to that of the first and second families of models and the decay time has no effect on the fractional rms amplitude.



\subsection{3C~390.3 \label{sec:app-3c390.3}}

Following the same procedure as for Arp~102B, we constructed the mean and average spectra from the 35 line profiles of 3C~390.3 and fitted both of these spectra with an axisymmetric power-law emissivity model of the form $\epsilon\propto\xi^{-3}$. The best fit results for the average profile are  $i=27^\circ\pm 2^\circ$, $\xi_{\rm min}=420^{+60}_{-50}$, $\xi_{\rm out}=1400^{+800}_{-250}$, $\sigma=1500\pm300$ km s$^{-1}$. The best fit parameters to the minimum spectrum are $i=27^\circ \pm 1^\circ$, $\xi_{\rm min}=500^{+100}_{-40}$, $\xi_{\rm out}=1500^{+1000}_{-300}$, $\sigma=1200^{+200}_{-400}$ km s$^{-1}$. The average spectrum and the best fit are shown in Figure \ref{fig:fig_av}b.

Thus, for the analysis of 3C~390.3 we set the disk parameters to the following values: $i = 27^\circ$, $\xi_{\rm min}=450$, $\xi_{\rm out}=1400$ and $\sigma=1300$ km s$^{-1}$. The mass of the black hole is set to $3\times 10^8 M_{\odot}$ (see Lewis \& Eracleous 2006). Hence, the dynamical times at the inner and outer radii are 0.45 and 2.5~years, respectively.



We explored the parameter space of the three families of models as we did for Arp~102B (\S \ref{sec:app-arp102b}) and we found similar correlations for all the families. The correlations of the power-law indices with the radial distribution of spots are not as strong as those obtained for Arp~102B. This is probably a result of the fact that 3C~390.3 has a larger disk size and naturally has more spots at large radii than Arp~102B.

\begin{figure}
\includegraphics[width=6in]{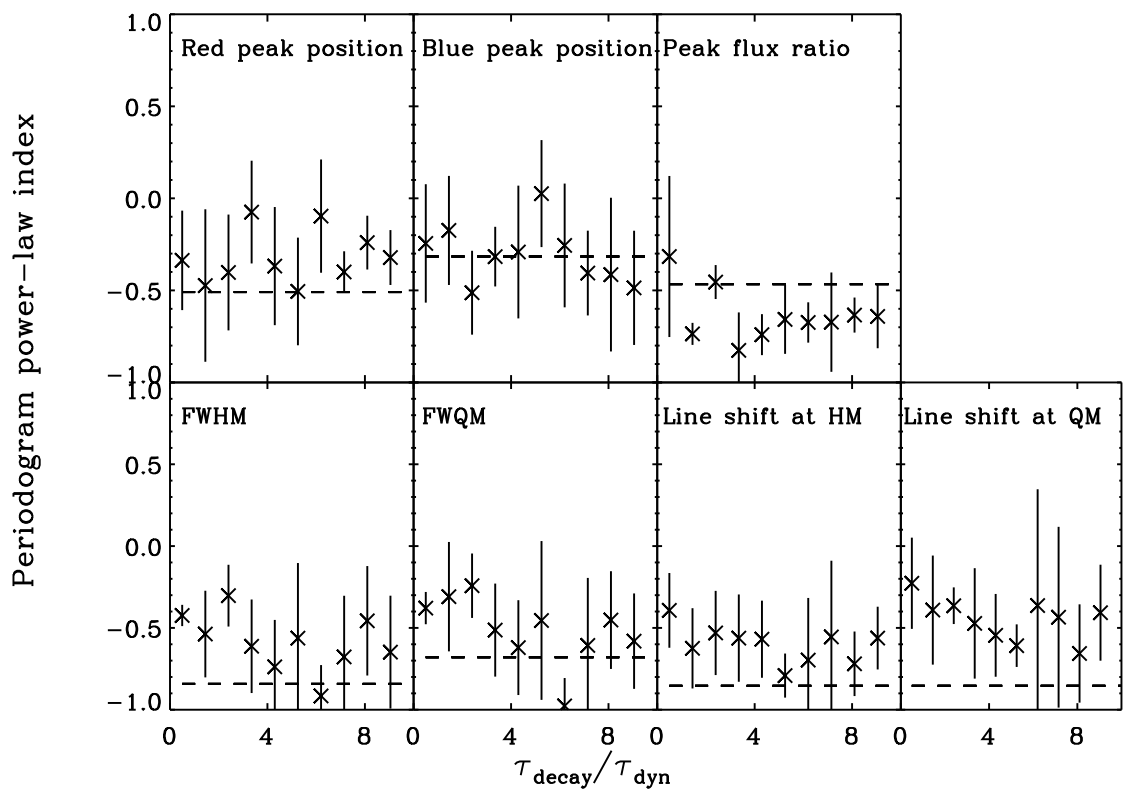}
\label{fig:match3c}
\caption{Examples of realization of the shearing spot model for which the power-law indices of the line parameter periodograms are close to the observed values for 3C~390.3. The ratio of the decay time to dynamical time is the only parameter that was allowed to vary. The horizontal dashed lines represent the value of the power-law index found in the corresponding observed periodogram of 3C~390.3. }
\end{figure}

Since the observed power-law indices of the periodohrams of 3C~390.3 are lower than those of Arp~102B, a wide range of model parameters can reproduce the values observed for 3C~390.3.  Figure \ref{fig:match3c} shows that the power-law indices of simulated periodograms are close to the observed values for 3C~390.3, here for a shearing spot model with a $\xi^2$ radial distribution of the spots and variable $\tau_{\rm decay}$. Such a close match is easily obtained for any family of models, depending on the initial parameter values.

The 2-D periodograms of 3C~390.3 have the same behavior and same typical values of the power-law indices as those of Arp~102B. Similarly, the fractional rms amplitude increases with increasing contrast and size of the spots, but is not sensitive to the other model parameters. A combination of large spot contrast and large spot size is required to match the observed fractional rms amplitide of 3C~390.3.



\section{Discussion}

Stochastically perturbed disk models produce variability of the H$\alpha$ line profile whose properties depend on the model parameters. Allowing for decay of the spots steepens the observed periodograms of the line parameters. This is because the decay introduces a new, longer timescale in the system, thus shifting power to lower frequencies The radial distribution of the spots also influences the power-law indices of the line parameter periodograms: steeper periodograms are produced when spots are concentrated towards the outer parts of the line-emitting region. This is a result of the fact that in such a distribution more spots have longer orbital periods, introducing more power at lower frequencies, thus steepening the periodogram. Shearing does not significantly change the power-law indices of the periodograms because shearing and decaying happen simultaneously and the effect of decaying of a spot is more prominent that that of shearing. Because shearing does not modify the variability properties of the line profile, one cannot constrainr it from observations of the line profile variability. The fractional rms amplitude of the line parameters increases with increasing contrast and size of the spots and does not depend strongly on the number of spots, their radial distribution, or their shear and decay properties. The fractional rms amplitude thus provides another constraint on the contrast of the perturbation of the emissivity function. 
 

The perturbed disk model can reproduce the observed variability properties of Arp~102B and 3C~390.3 with different levels of success. Most of the observed line parameter periodograms of Arp~102B are too steep to be successfully reproduced. The best models have decaying spots, mostly concentrated towards the outer parts of the line emitting region. Such models produce power-law indices of the peak flux ratio and blue peak position periodograms that are comparable to those of Arp~102B, but all the other line parameter power-law indices are too high by up to 0.4. The maximum radial spot distribution index that we tested is +2. A model with an even steeper radial distribution might produce steep enough periodograms to match the observations. Such a distribution rules out the production of spots by collisions with stars since this process skews the spot distribution towards the inner part of the accretion disk. Spot production by vortices yields a priori a uniform spot distribution. Spot production by self gravity leads to spots only past the self gravity radius so this could skew the spot distribution towards the outer part of the line emitting region. These spots would however neither shear nor decay so one would need an even steeper radial distribution index to achieve agreement with the observed values of Arp~102B. A sharp increase from 0 to a large number of spots past the radius of marginal self-gravity might mimic such a steep radial distribution index. Using the properties of Arp~102B determined by Lewis \& Eracleous(2006) and the formulae for the radius of marginal self-gravity ($r_{\rm sg}$) given by Goodman \& Tan (2004) and Hur\'e (1998), we estimate that $400<r_{\rm sg}/r_{\rm g}<900$ by varying the disk parameters within their plausible range. This estimate falls within the range of the line emitting region determined by fitting the average spectrum. To test whether a sharp boundary in the spot distribution steepens the power-law indices of the periodograms, we ran one simulation of the first family of models with no spots below 750 $r_{\rm g}$ and 100 spots uniformly distributed above that radius. The spots were large (15 $r_{\rm g}$) and bright ($\langle C_i\rangle =20$) as one would expect from self-gravitating clumps. The resulting line parameter periodograms had power-law indices varying between $-0.17$ and $-0.72$, which is steeper than what a smooth radial spot distribution produces and closer to the values observed for Arp~102B. Moreover, these self-gravitating spots would have the necessary size and contrast to match the observed fractional rms amplitude of Arp~102B. This result bolsters the hypothesis that spots inhabit the outer parts of the line-emitting disk of Arp~102B.

There could also be other factors that steepen the line parameter periodograms. We assumed constant inner and outer radii of the line emitting region, but we tested the effect of a {\it fast} modulation of these radiii and found that this produces steeper periodograms. The inner and outer radii could vary due a change in the disk ionization structure after a variation of the ionizing continuum from the central source. Developing a physically self-consistent model to account for this effect requires one to calculate the variation of  the whole disk structure with ionizing flux and is beyond the scope of this paper.

The line parameter periodograms of 3C~390.3 are not as steep as those of Arp~120B and can be repoduced by a wide range of model parameters. A radial distribution of spots skewed towards the outer parts of the line emitting region is not absolutely necessary for 3C~390.3 for two reasons. First, the steepening of the line parameter periodograms with increasing radial spot distribution index is not as marked as for Arp~102B. This might be a result of the fact that the line emitting region of 3C~390.3 has a much larger outer radius than that of Arp~102B. Hence, there is naturally a larger number of spots at large radii. This suggests that it is the relative number of spots at large radii compared to lower radii rather than the absolute number of spots at large radii that influences the power-law indices of the periodograms. Also since the observed power-law indices of the line parameter periodograms of 3C~390.3 are higher than those of Arp~102B, we do not need to select the model families and parameters producing steep power spectra in order to reproduce the properties of 3C~390.3. This also means that no particular model for the origin of spots in the accretion disk of 3C~390.3 can be ruled out or favored. We note that $250<r_{\rm sg}/r_{\rm g}<400$ for 3C~390.3, which is below $\xi_{\rm min}$ and strengthens our assumption of a perturbed accretion disk at all radii. 

The behavior of the collapsed periodograms from the 2-D periodogram is a greater challenge to this model. The difference between the two power-law indices agree with the observed difference, but the values of the indices are too high to be reconciled with the observations.

The results of our line profile variability study are quite different between Arp~102B and 3C~390.3 even though their black hole masses are comparable. The main driver of this difference is the higher accretion rate of 3C~390.3 ($0.15~{\rm M}_\odot~{\rm yr}^{-1}$ vs. $1.7\times 10^{-3}~{\rm M}_\odot~{\rm yr}^{-1}$ for Arp~102B), which makes that disk more unstable to self-gravity. It is encouraging that the perturbed accretion disk model provides a unified solution to explain the profile variability of two AGNs that are so different. In the context of the production of clumps by self-gravity, AGNs with similar black hole masses and even higher Eddington ratios than 3C~390.3 are expected to be unstable at all radii and the variability of their profile would then be well explained by our model. AGNs with smaller black hole masses might have their radius of marginal self-gravity in the line-emitting region and have a variability pattern similar to Arp~102B. In order to test this theory and confirm our result, we need to apply this method to a larger sample of AGNs covering a wider range of black hole masses and accretion rates, such as those presented in Gezari et al. (2007) and Lewis (2005).  

\acknowledgements
This work was partially supported by the Zaccheus Daniel grant. We would like to thank Suvi Gezari for providing fully reduced spectra for Arp~102B and 3C~390.3. We also want to thank Phil Uttley for a helpful discussion on periodograms and Fourier transforms. Finally we want to thank the anonymous referee for thoyghtful comments and suggestions.

\clearpage

\begin{table}[h]
\begin{center}
\caption{Observed fractional rms amplitude of the line profile parameters of Arp~102B and 3C~390.3}
\begin{tabular}{lcc}
\hline \hline
&	Arp~102B	& 3C~390.3 \\
\hline
Red Peak Position (km~s$^{-1}$)	&	$(8.9 \pm 0.3)\times 10^{-2}$	&	0.35 $\pm$ 0.03\\
Blue Peak Position (km~s$^{-1}$)	& 	$(4.4 \pm 0.4) \times 10^{-2}$ &$(6.7 \pm 0.1)\times 10^{-2}$\\
Peak Flux ratio	&	$0.10 \pm 0.02$	&	0.15 $\pm$ 0.02	\\
FWHM (km~s$^{-1}$)	&	$(3.9 \pm 0.2) \times 10^{-2}$ & $(3.8 \pm 0.03)\times 10^{-2}$ \\
FWQM (km~s$^{-1}$)	& $ (4.4 \pm 0.2)\times 10^{-2}$ & $(4.3 \pm 0.02)\times 10^{-2}$\\
Displacement at HM (km~s$^{-1}$)	&	0.40	$\pm$ 0.04&	4.6 $\pm$ 0.6\\
Displacement at QM (km~s$^{-1}$)	&  0.31 $\pm$ 0.03& 1.0 $\pm$ 0.1\\
\hline
 \end{tabular}
 \label{tab:rms}
\end{center}

\end{table}


\begin{thebibliography}{}
\bibitem[Abramowicz et al. 1991]{Abram91}Abramowicz, M.A., Bao, G., Lanza, A., \& Zhang, X.-H. 1991, A\&A, 245, 454
\bibitem[Bao \& {\O}stgaard 1995]{BO95}Bao, G., \& {\O}stgaard, E. 1995, ApJ, 443, 54
\bibitem[Barranco \& Marcus 2005]{BM05}Barranco, J.A., \& Marcus, P.S. 2005, ApJ, 623, 1157
\bibitem[Chen \& Halpern 1989]{CH89} Chen, K., \& Halpern, J.P. 1989, ApJ, 344, 115
\bibitem[Chen et al. 1989]{Chenal89} Chen, K., Halpern, J.P., \& Filippenko, A.V. 1989, ApJ, 339, 742
\bibitem[Collin-Souffrin 1987]{CS87}Collin-Souffrin, S. 1987, A\&A, 179, 60
\bibitem[Collin-Souffrin \& Dumont 1990]{CSD90} Collin-Souffrin, S., \& Dumont, A.M. 1990, A\&A, 229, 292
\bibitem[Dietrich et al. 1998]{Dietal98} Dietrich, M., et al. 1998, ApJS, 115, 185
\bibitem[Dumont \& Collin-Souffrin 1990] {DCS90} Dumont, A.M., \& Collin-Souffrin, S. 1990, A\&A, 229, 302
\bibitem[Eracleous \& Halpern 1994]{EH94} Eracleous, M. \& Halpern, J. P. 1994, ApJS, 90, 1
\bibitem[Eracleous \& Halpern 2003]{EH03} Eracleous, M. \& Halpern, J. P. 2003, ApJ, 599, 886
\bibitem[Eracleous et al. 1995]{Eracal95} Eracleous,  M., Livio, M., Halpern, J. P., \& Storchi-Bergmann, T. 1995, ApJ, 438, 610
\bibitem[Eracleous et al. 1997]{Eracal97}Eracleous, M., Halpern, J. P., Gilbert, A. M., Newman, J. A., \&
Filippenko, A. V. 1997, ApJ, 490, 216
\bibitem[Gaskell 1983]{Gskell83}Gaskell, C.M. 1983, in Proc 24th Li\`ege Int. Astrophys. Colloq., Quasars and Gravitational Lenses (Cointe-Ougree: Univ. Lie`ge), 473 
\bibitem[Gezari et al. 2007]{Gezarith}Gezari, S., Halpern, J.P., \& Eracleous, M. 2007, ApJ, 169, 167
\bibitem[Gilbert et al. 1999]{Gilbert}Gilbert, A.M., Eracleous, M., Filippenko, A.V., \& Halpern, J.P. 1999, ASPC, 175, 189 
\bibitem[Goodman \& Tan 2004]{GT04}Goodman, J., \& Tan, J.C. 2004, ApJ, 608, 108
\bibitem[Hur\'e 1998]{H98}Hur\'e, J.-M. 1998, A\&A, 337, 625
\bibitem[Kassebaum et al. 1997]{Kassebaum97}Kassebaum, T.M., Peterson, B.M., Wanders, I., Pogge, R.W., Bertram, R., \& Wagner, R.M. 1997, ApJ, 475, 106
\bibitem[Lawrence \& Papadakis 1993]{LP93}Lawrence, A., \& Papadakis, I. 1993, ApJ, 414, 85
\bibitem[Lewis 2005]{Lewis05}Lewis, K.T. 2005, PhD Thesis, The Pennsylvania State University
\bibitem[Lewis \& Eracleous 2006]{LE06}Lewis, K.T., \& Eracleous, M. 2006, ApJ, 642, 711
\bibitem[Marziani et al. 1993]{Marziani93}Marziani, P., Sulentic, J. W., Calvani, M., Perez, E., Moles, M., \& Penston, M. V. 1993, ApJ, 410, 56
\bibitem[Murray \& Chiang 1997]{MC97}Murray, N., \& Chiang, J. 1997, ApJ, 474, 91
\bibitem[Newman et al. 1997]{Newman97}Newman, J. A., Eracleous, M., Filippenko, A. V., \& Halpern, J. P. 1997, ApJ, 485, 570
\bibitem[OÕBrien et al. 1998]{Obrien98}OÕBrien, P. T., et al. 1998, ApJ, 509, 163
\bibitem[Papadakis \& Lawrence 1993]{PL93}Papadakis, I., \& Lawrence, A. 1993, Nature, 361, 233
\bibitem[Pariev \& bromley 1998]{PB98}Pariev, V.I., \& Bromly, B.C. 1998, ApJ, 508, 590
\bibitem[Petersen et al. 2007]{PSJ07}Peterson, M.R., Stewart, G.R., \& Julien, K. 2007, ApJ, 658, 1252
\bibitem[Peterson et al. 1982]{Petersonal82}Peterson, B.M., Foltz, C.B., Byard, P.L., \& Wagner, R.M., 1982, ApJS, 49, 469 
\bibitem[Press \& Rybicki 1989]{PR89} Press, W.H., \& Rybicki, G.B. 1989, ApJ, 338, 277
\bibitem[Rokaki et al. 1992]{Rokakietal92}Rokaki, E., Boisson, C, \& Collin-Souffrin, S. 1992, A\&A, 253, 57
\bibitem[Romano et al. 1998]{Romano98}Romano, P., Marziani, P., \& Dultzin-Hacyan, D. 1998, ApJ, 495, 222
\bibitem[Rosenblatt et al. 1992]{Rosenetal92}Rosenblatt, E.I., Malkan, M.A., Sargent, W.L.W., \& Redhead, A.C.S. 1992, ApJS, 81, 59
\bibitem[Scargle 1982]{Scargle82}Scargle, J.D. 1982, ApJ, 263, 835
\bibitem[Sergeev et al. 2000]{Sal00}Sergeev, S.G., Pronik, V.I., \& Sergeeva, E.A. 2000, A\&A, 356, 41
\bibitem[Shakura \& Sunyaev 1973]{SS73}Shakura, N.I., \& Sunyaev, R.A. 1973, A\&A, 24, 337
\bibitem[Shapovalova et al. 2001]{Shap01} Shapovalova, A.I., Burenkov, A. N., Carrasco, L., Chavushyan, V. H., Doroshenko, V. T., Dumont, A. M., Lyuty, V. M., ValdŽs, J. R., Vlasuyk, V. V., Bochkarev, N. G., Collin, S., Legrand, F., Mikhailov, V. P., Spiridonova, O. I., Kurtanidze, O., \& Nikolashvili, M. G. 2001, A\&A, 376, 775
\bibitem[Storchi-Bergmann et al. 2003]{SBal95} Storchi-Bergmann, T., Nemmen da Silva, R., Eracleous, M., Halpern, J. P., Wilson, A. S., Filippenko, A. V., Ruiz, M. T., Smith, R. C., Nagar, N. 2003, ApJ, 598, 956
\bibitem[Storchi-Bergmann et al. 2003]{Storchietal03} Storchi-Bergmann, T., Nemmen da Silva, R., Eracleous, M., Halpern, J.P., Wilson, A.S.,Filippenko, A.V., Ruiz, M.T., Smith, R.C., \& Nagar, N.M. 2003, ApJ, 598, 956 
\bibitem[Strateva et al. 2003]{Sal03}Strateva, I. V. et al. 2003, AJ, 126, 1720
\bibitem[Turner et al. 2005]{Turner05}Turner, T.J., Miller, L., George, I.M., \& Reeves, J.N. 2005, A\&A, 445, 59
\bibitem[Veilleux \& Zheng 1991]{VZ91}Veilleux, S., \& Zheng, W. 1991, ApJ, 377, 89 
\bibitem[Wanders et al. 1995]{Wanders95}Wanders, I., et al. 1995, ApJ, 453, L8
\bibitem[Wanders \& Peterson 1997]{WP97}Wanders, I., \& Peterson, B.M. 1997, ApJ, 477, 990
\bibitem[Zheng et al. 1990]{Zheng90}Zheng, W., Binnette, L., Sulentic, J.W. 1990, ApJ, 365, 115
\bibitem[Zheng et al. 1991]{Zheng91}Zheng, W., Veilleux, S., \& Grandi, S.A. 1991, ApJ, 381, 418
\bibitem[Zurek et al. 1995]{Zurekal95}Zurek, W.H., Siemiginowska, A., \& Colgate, S.A. 1995, ApJ, 434, 46
\end{thebibliography}
\end{document}